\documentclass[pdftex,epjc3]{svjour3}  
\usepackage{geometry}        
\usepackage{amsmath,amssymb}
\usepackage{comment,mfirstuc}
\usepackage[utf8]{inputenc}
\usepackage[T1]{fontenc}
\usepackage{booktabs}
\usepackage{subcaption}

\newcommand{\addComment}[2]{
  \expandafter\newcommand\csname #1\endcsname[1]{{\bf \color{#2} \capitalisewords{#1}:\,##1}}
  \expandafter\newcommand\csname #1cor\endcsname[2]{{\color{#2} \capitalisewords{#1}:\,\st{##1}{\bf ##2}}}
  \expandafter\newcommand\csname #1color\endcsname{#2}
}
\addComment{cris}{blue} 
\addComment{james}{red}
\addComment{alex}{orange}

\RequirePackage[T1]{fontenc}

\smartqed  

\RequirePackage{graphicx}
\RequirePackage{mathptmx}      
\RequirePackage{flushend}
\RequirePackage[numbers,sort&compress]{natbib}
\RequirePackage[colorlinks,citecolor=blue,urlcolor=blue,linkcolor=blue]{hyperref}

\journalname{EPJ Data Science}

\begin{document}

\title{Unmasking Social Bots: How Confident Are We?} 


\renewcommand{\and}{\hspace{10pt}} 


\author{\mbox{James Giroux\thanksref{addr1}, 
              Gangani Ariyarathne\thanksref{addr1}, 
              Alexander C. Nwala\thanksref{addr1,addr2},        
              Cristiano Fanelli\thanksref{addr1}}
              \texttt{\color{blue}\{jgiroux, gchewababarand, acnwala, cfanelli\}@wm.edu} \\
}




\institute{
\raggedright William \& Mary, Department of Data Science, Williamsburg, VA 23185, USA\label{addr1}
\and 
\raggedright \ Observatory on Social Media, Indiana University, Bloomington, IN 47405, USA\label{addr2}
}

\date{}

\maketitle

\begin{abstract}
Social bots remain a major vector for spreading disinformation on social media and a menace to the public. Despite the progress made in developing multiple sophisticated social bot detection algorithms and tools, bot detection remains a challenging, unsolved problem that is fraught with uncertainty due to the heterogeneity of bot behaviors, training data, and detection algorithms. Detection models often disagree on whether to label the same account as bot or human-controlled. However, they do not provide any measure of uncertainty to indicate how much we should trust their results. 
We propose to address both bot detection and the quantification of uncertainty at the account level --- a novel feature of this research. This dual focus is crucial as it allows us to leverage additional information related to the quantified uncertainty of each prediction, thereby enhancing decision-making and improving the reliability of bot classifications. Specifically, our approach facilitates targeted interventions for bots when predictions are made with high confidence and suggests caution (\textit{e.g.}, gathering more data) when predictions are uncertain.
\end{abstract}

\keywords{Uncertainty quantification, Bayesian neural network, Social media, Bot detection}

\section{Introduction}

Social media platforms have fundamentally transformed global communication, enabling the near-instant dissemination of information. Platforms like Facebook, YouTube, and Twitter/X, have over two billion monthly active users~\cite{social_media_user_stats} and empower individuals to broadcast their thoughts easily. However, the popularity that social media enjoys has incentivized malicious actors such as tech-savvy individuals or governments~\cite{twitter_info_ops}, to deploy social bots to influence organic discourse and manipulate social media users for economic or political profit. Social bots~\cite{ferrara2016rise,cresci2020decade} --- accounts controlled partly or fully by software --- have been used to artificially boost the popularity of political candidates~\cite{ratkiewicz2011detecting}, spread conspiracy theories~\cite{Lazer-fake-news-2018,grinberg2019fake} during health crises~\cite{himelein2021bots}, and to manipulate the stock market~\cite{cresci2019cashtag,pacheco2020uncovering}. 

\textcolor{black}{According to} a 2023 study that analyzed 1 million tweets during a Republican debate and a Donald Trump interview,  over 1,200 bot accounts were identified as spreading false narratives, highlighting the significant impact of bots during major events \cite{graham2023bots}. Anecdotal reports further underscore the prevalence of bot accounts, with some users suggesting that a substantial fraction of interactions on the platform are bot-generated. Despite enhanced verification processes, verified accounts continue to promote fraudulent schemes, illustrating the ongoing challenge in curbing bot activity.

\begin{figure}[!]
    \centering
    \includegraphics[width=0.90\textwidth]{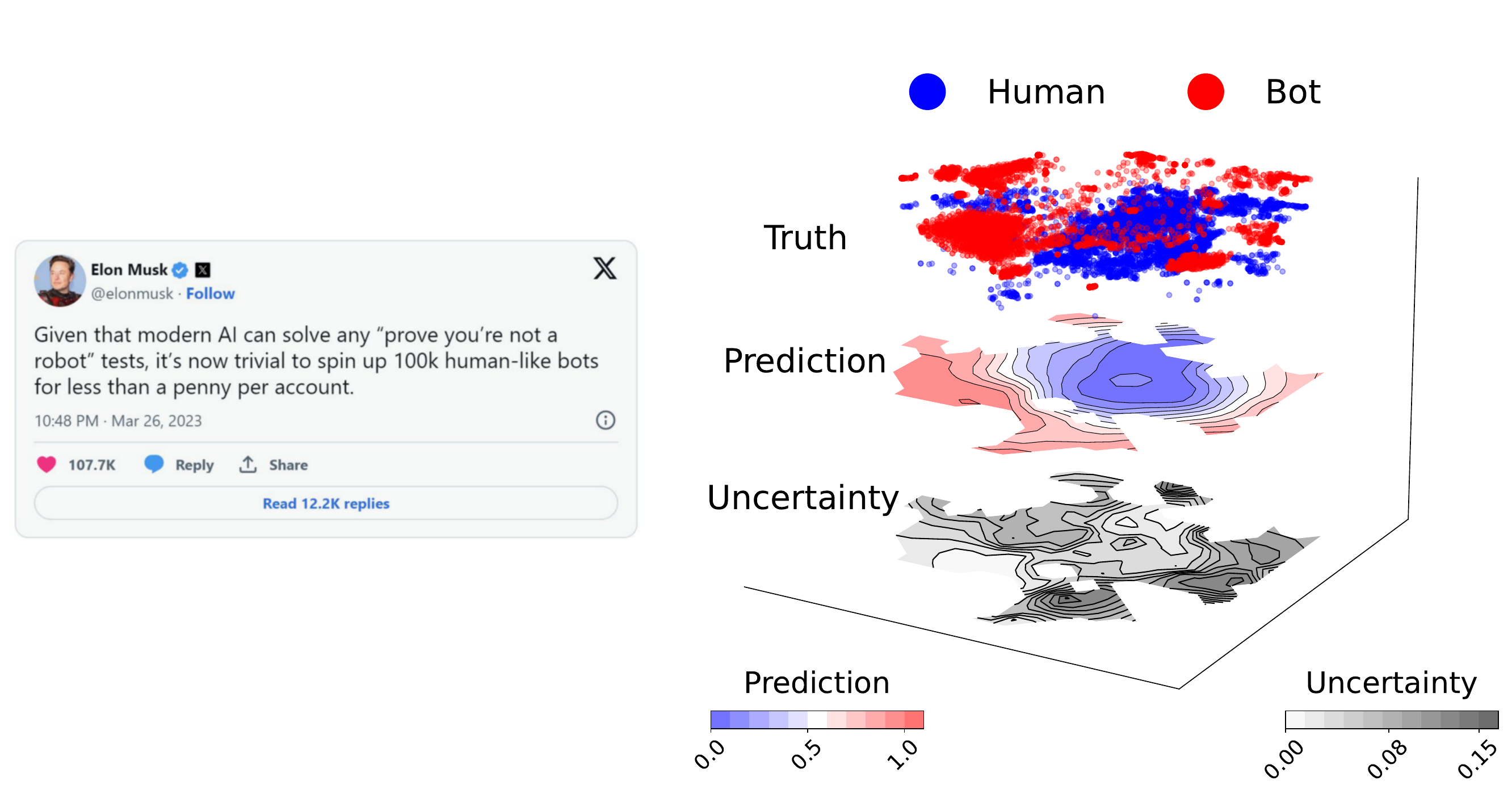}
    \caption{\textbf{Impact of Bot Crisis and Uncertainty on Predictions (Our Work):}
    As stated by Musk through X (left), the ability of bots to replicate human behavior and bypass security measures has increased dramatically with the advent of AI. Bot accounts are able to more efficiently mask themselves within the human population on social media platforms. This is shown through dimensionally reduced representations (right), in which we show three planes: (i) the true distributions, where we introduce an offset between human and bot points to ease visualization; (ii) the expected probability of an account being a bot as produced by our network, along with (iii) the associated uncertainty across the feature space, represented by the epistemic and aleatoric components added in quadrature. For (ii) and (iii) we use Gaussian process regression \cite{rasmussen2003gaussian} for visualization purposes. Uncertainty is greater in regions where ambiguity is higher and the two classes overlap. 
    }
    \label{fig:3d_tsne}
\end{figure}

The issue of bot prevalence has also featured prominently in high-profile disputes, such as Elon Musk's acquisition of Twitter. Musk's legal team, using Botometer \cite{yang2022botometer}, an online tool for identifying spam and fake accounts, claimed that 33\% of ``visible accounts'' were ``false or spam''. However, Botometer's creator, Kaicheng Yang, criticized this claim and their methodology, stating that the figure was misleading and disclosed that Musk's team had not consulted him before using the tool \cite{bbctech2023}. Perhaps even more concerning is the adaptation of Artificial Intelligence (AI) to bypass security systems  deemed ``prove you're not a robot tests.'' This concern has been echoed by Musk, Fig. \ref{fig:3d_tsne} (left), as these tests are commonly used as initial filters for the prevention of inauthentic and/or malicious bot accounts. Moreover, accounts that successfully bypass such tests are increasingly becoming more ``human-like,'' as indicated by the dimensionality reduced representations of accounts in Fig. \ref{fig:3d_tsne} (right). 
\textcolor{black}{The figure illustrates three levels: the top represents the ground truth distribution, where many bots appear human-like. The middle plane shows classifier predictions, with blue favoring humans, red for bots, and white indicating indecision. The bottom plane depicts prediction uncertainty, with darker regions corresponding to higher uncertainty, aligning with indecisive areas in the middle plane.}
%
Musk has \textcolor{black}{suggested} that implementing a paid subscription model could be an effective way to combat bots on X (formerly Twitter).
%
%
%
%
\textcolor{black}{While} the introduction of a paid subscription model aimed to enhance verification, \textcolor{black}{concerns remain} that it may not have fully addressed the issue of inauthentic accounts. Some argue that this approach could allow certain accounts to obtain verification status (blue check mark) through payment, potentially lending them an appearance of authenticity.
This issue has been highlighted by the European Commission, which has claimed that Twitter was in violation of the Digital Services Act~\cite{2024_ec_twitter_dsa_breach}.
%
%
Similarly, the \textcolor{black}{decision} to introduce a high-cost paywall for access to Twitter's research Application Programming Interface (API) has \textcolor{black}{posed challenges} for researchers in studying and addressing bot activity.

Given the serious harm that social bots wielded by bad actors, pose to democracy~\cite{schiffrin2017disinformation, woolley2018computational}, public health~\cite{tasnim2020impact, allington2020health, covaxxy-misinfo}, or the the economy~\cite{fisher2013syrian}, researchers have responded by developing a broad range of bot detection tools. Various Machine-Learning methods have been trained to detect social bots with a combination of features extracted from the social network structure, content/profile attributes, and temporal patterns~\cite{ferrara2016rise}. Alternatively, all of these feature types are combined into a single model~\cite{lee2011seven,davis2016botornot,varol2017online,gilani2017bots,yang2019arming,sayyadiharikandeh2020detection}. Bot-detection algorithms often start by modeling the characteristics of accounts as the first step to distinguish bot from human-like behaviors~\cite{cresci2020decade}. Accounts may be represented using user profile information~\cite{yang2020scalable}, content~\cite{yardi2010detecting,cresci2017social,cresci2017exploiting}, actions~\cite{mazza2019rtbust}, social network~\cite{beskow2018bot}, and temporal signatures~\cite{mazza2019rtbust}.

Despite the progress made, bot detection remains a challenging, unsolved problem that is fraught with uncertainty due to the heterogeneity of bot behaviors, training data, and detection algorithms~\cite{cresci2023demystifying}. Consequently, it is not surprising for detection models to disagree on whether to label the same account as bot or human, since they are often trained on different datasets that are sensitive to different subsets of the signals of automation. In light of this, it is paramount that bot detection models provide uncertainty weights alongside the probability estimate that an account is bot-controlled, as this could determine if one should trust a prediction. However, existing bot detection algorithms focus exclusively on detection. 

Our work addresses both bot detection and the quantification of uncertainty at the account level, an important distinctive feature of this research. This dual focus is crucial as it allows us to leverage additional information related to the quantified uncertainty of each prediction, thereby enhancing decision-making and improving the reliability of bot classifications. Specifically, our approach facilitates targeted interventions for bots where predictions are made with high confidence and suggests a cautious approach, \textit{e.g.}, gathering more data for bots where predictions are uncertain. Also, our method is agnostic to bot-detection algorithm, which enables the inclusion of uncertainty measurement into existing bot detection systems. As discussed earlier, research at the nexus of uncertainty quantification and bot detection is both timely and novel. In subsequent sections, we present our methodology that adeptly separates aleatoric and epistemic uncertainties. Aleatoric uncertainty, arising from the inherent randomness in our bot datasets, is characterized using multiple features from the Behavioral Language for Online
Classification (BLOC) framework \cite{nwala2023language}, or features from Botometer \cite{yang2022botometer}. Conversely, epistemic uncertainty originates from the limitations of the predictive model used.
Previous studies such as \cite{taheri2022unbus} have employed Deep Ensemble and Stochastic Weight Averaging (SWA) for uncertainty estimation, yet these methods have notable limitations \cite{abdar2021review}. SWA does not differentiate between the two types of uncertainties and requires supplementary techniques like dropout or bootstrapping for effective uncertainty quantification. Similarly, while Deep Ensembles capture epistemic uncertainty, they fail to distinctly separate it from aleatoric uncertainty.

In contrast, our approach utilizes Bayesian methods, which provide a comprehensive and theoretically grounded framework for distinguishing and quantifying different types of uncertainties. We demonstrate that the computational demand of our Bayesian Neural Network (BNN) is negligible given the complexity of the problem at hand, and agnostic to input, \textit{e.g.} BLOC or Botometer features. This approach positions our methodology as exceptionally capable in the critical domain of bot detection, offering a robust mechanism for managing uncertainty quantification.



\textcolor{black}{In Sec. \ref{subsec:bot_det} we provide a literature review of bot detection methods, along with Uncertainty Quantification within the scope of bot detection in Sec. \ref{subsec:uq_bot}. In Sec. \ref{sec:methods} we provide detailed descriptions of the feature extraction methods (BLOC and Botometer), and the Bayesian Deep Learning methods used. Sec. \ref{sec:exps} details the experimental setup in terms of datasets, training and inference procedures, followed by Sec. \ref{sec:results} in which we present our results and performance metrics obtained through the implementation of these methods. Finally, Sec. \ref{sec:conclusions} offers a summary of our findings and conclusions. The contributions of our work are as follows:}

\begin{itemize}
    \item We introduce the first fully Bayesian Deep Learning method of uncertainty quantification in the space of bot detection; capable of providing both epistemic (model) and aleatoric (stochastic) uncertainties.
    \item Our method provides uncertainty-aware decisions at the account level. Capable of improving performance of the architecture over baselines.
    \item Our method is agnostic to bot detection features, implying usage in downstream tasks from other pre-processing schemes or Deep Learning based feature extraction algorithms.
\end{itemize}

\section{Related Works}\label{sec:related_works}

\subsection{Bot Detection}\label{subsec:bot_det}
Social media abusers utilize different tactics to manipulate their audiences for political~\cite{woolley2018computational} or economic~\cite{fisher2013syrian} gain. One of of the earliest forms of abuse was spamming by spam bots (software-controlled accounts). Spam accounts were easy to detect because they lacked meaningful profile information and/or demonstrated naive behaviors~\cite{yardi2010detecting,lee2011seven}. However, following the development of effective spam account detection methods (e.g.,~\cite{markines2009social,jin2011socialspamguard,ZHENG201527, stringhini2010detecting,khan2015comprehensive,rastogi2020effective,rastogi2017opinion}), spam bots evolved to social bot accounts~\cite{ferrara2016rise, cresci2020decade}. Similar to spam bots, social bots are controlled fully or partly by software, but are more sophisticated. For example, some accounts have detailed profiles, either stolen from real users or generated by deep neural networks~\cite{nightingale2022ai}. Some can interact with actual humans or mimic human behaviors by generating human-like content with ChatGPT~\cite{Yang2023Anatomy-AI-botnet} and build social connections~\cite{cresci2020decade}. Others like cyborgs~\cite{chu2012detecting, chu2010tweeting} cycle between human and bot-like behaviors.

Various Machine-Learning methods have been trained to detect social bots with a combination of features extracted from the social network structure, content/profile attributes, and temporal patterns~\cite{ferrara2016rise}. Alternatively, all of these feature types are combined into a single model~\cite{lee2011seven,davis2016botornot,varol2017online,gilani2017bots,yang2019arming,sayyadiharikandeh2020detection,cai2024lmbot}. Bot-detection algorithms often start by modeling the characteristics of accounts as the first step to distinguish bot from human-like behaviors~\cite{cresci2020decade}. Accounts may be represented using user profile information~\cite{yang2020scalable}, content~\cite{yardi2010detecting,cresci2017social,cresci2017exploiting}, actions~\cite{mazza2019rtbust}, social network~\cite{beskow2018bot}, and temporal signatures~\cite{mazza2019rtbust}.

The algorithms described here only produce probabilities estimating the likelihood that accounts are bots (software controlled). Since bot detection remains a challenging unsolved problem due to the heterogeneity of bot behaviors, training data, and detection algorithms~\cite{cresci2023demystifying}, different detection methods could disagree on the label to assign to the same account. Consequently, an important novel contribution of this work is addressing both bot detection and the quantification of uncertainty at the account level. This enables us to leverage additional information related to the quantified uncertainty of each prediction, thereby enhancing decision-making and improving the reliability of bot classifications.


\subsection{Uncertainty Quantification in Bot Detection}\label{subsec:uq_bot}

Uncertainty Quantification (UQ) is an increasingly critical component of decision-making, particularly with the adoption of AI-centric pipelines in the domain of bot detection. These pipelines often lack transparency due to their black-box nature and are typically unable to provide introspective confidence measures during inference. Consequently, UQ has emerged as a powerful tool, enabling fine-grained decision-making and identifying regions of low confidence (unreliability) within a model's output space.

A natural approach to UQ involves Bayesian methods, which enable the decomposition of uncertainty into two distinct sources: epistemic and aleatoric. Epistemic uncertainty arises from limitations in the model itself, while aleatoric uncertainty reflects the inherent randomness in the data. Despite the critical importance of uncertainty-informed decision-making in managing interactions on social platforms, there has been limited exploration of UQ in the context of bot detection, particularly within the space of modern deep learning architectures.

Existing methods such as Stochastic Weight Averaging (SWA) \cite{taheri2022unbus} are constrained by their reliance on approximating the posterior distribution through stochastic weight updates during the final training iterations. These approaches typically require additional techniques to produce meaningful uncertainty estimates. Similarly, Naive Bayes methods \cite{gamallo2019naive,kirn2021bayesian} often fall short due to their simplistic assumptions, rendering them inadequate for capturing both epistemic and aleatoric uncertainties.

In contrast, we present the first Bayesian deep learning framework for bot detection, which effectively separates and assesses both epistemic and aleatoric uncertainties. Our approach leverages state-of-the-art Bayesian deep learning techniques, inspired by advancements in Computer Vision (CV) \cite{kendall2017uncertainties}, alongside Multiplicative Normalizing Flows (MNF) \cite{pmlr-v70-louizos17a}, to model complex posterior distributions. Our network design philosophy is inherited from those developed in Nuclear Physics \cite{fanelli2024eluquant}. A detailed description of the implemented methodology is provided in Sec.~\ref{subsec:bnn}.

\section{Methods}\label{sec:methods}

In Sec. \ref{subsec:features} we describe the feature extraction methods used to form inputs to our bot detection pipelines, namely BLOC \cite{nwala2023language} and Botometer \cite{yang2022botometer}. We will then describe the inner working of our Bayesian approach in Sec. \ref{subsec:bnn}.

\subsection{\textbf{Feature Extraction}}\label{subsec:features}

\paragraph{\textbf{BLOC}}
\begin{figure}
  \centering
 \includegraphics[width=0.975\linewidth]{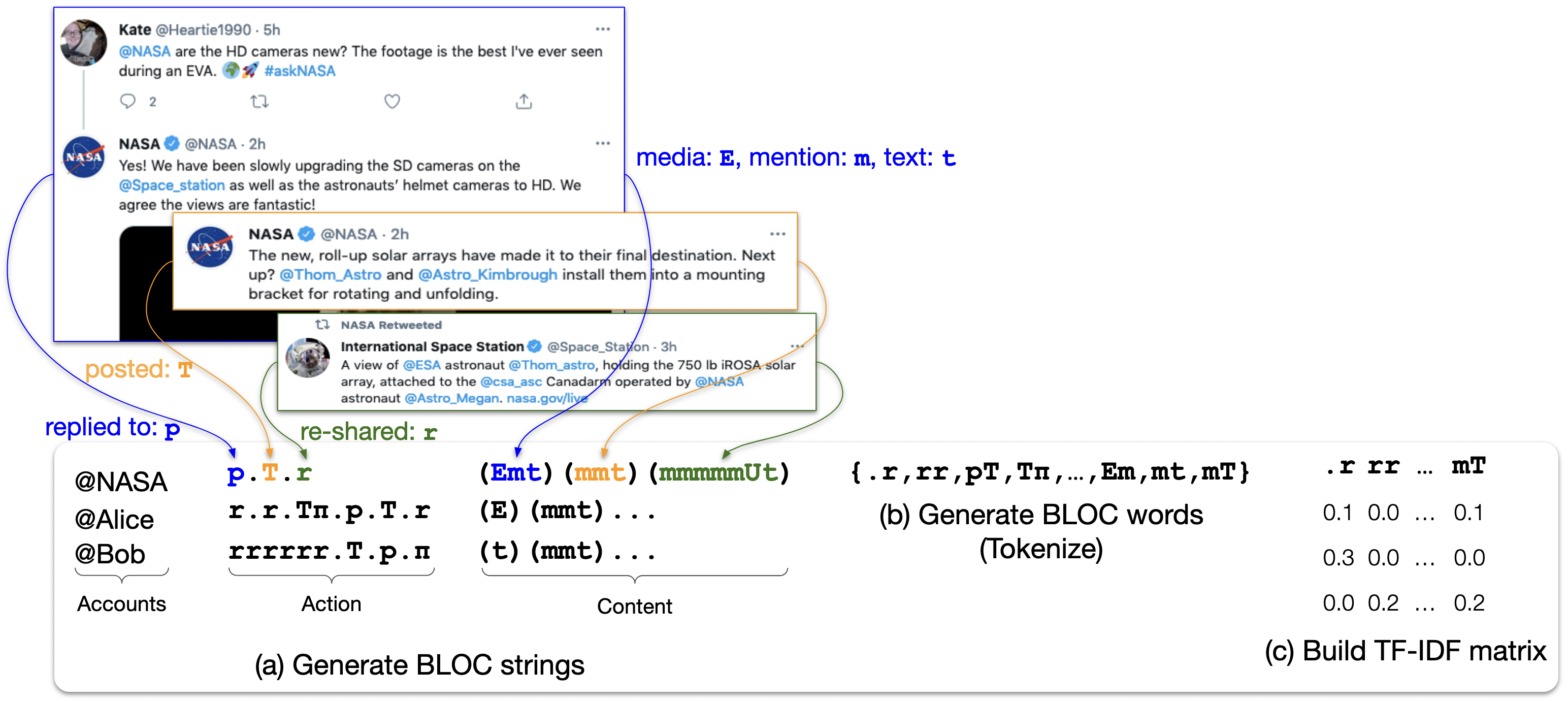}
 \caption{\textbf{BLOC Process Summary:}
 (a) BLOC action and content strings for three users, \texttt{@NASA}, \texttt{@Alice}, and \texttt{@Bob}. Using the action alphabet, the sequence of three tweets (a reply, an original tweet, and a retweet) by \texttt{@NASA} can be represented by three letters $p.T.r$ separated by dots (long pauses). Using the content alphabet, it can be represented by these sets of strings $(Emt)(mmt)(mmmmmUt)$ enclosed in parentheses. (b) After generating BLOC strings, they can be tokenized to generate words which are subsequently used to, (c) generate a matrix which serves as input to BNN and DNN.
 }
  \label{fig:bloc_process}
\end{figure}

The BLOC framework~\cite{nwala2023language} provides formal languages that represent the behaviors of social media accounts irrespective of social media platform, user-agent (human or bot), or intent (malicious or benign). The BLOC formal languages are defined by a set of alphabets (\textit{action} and \textit{content}) and rules for generating BLOC strings which are tokenized to produce BLOC words. BLOC words which represent the behaviors of social media accounts, consist of symbols drawn from distinct alphabets representing an account’s actions and content.  Fig.~\ref{fig:bloc_process}a illustrates a possible representation of a sequence of tweets by three different Twitter accounts, \texttt{@NASA}, \texttt{@Alice}, and \texttt{@Bob}. The  \texttt{@NASA} account replied to a tweet, posted a tweet, and then re-shared (retweet) a tweet, resulting in the BLOC action sequence: $p.T.r$. Here, each action (\textit{e.g.}, reply to) is represented by a single symbol (\textit{e.g.}, $p$), and the dots represent long pauses (\textit{e.g.}, $>$ 1 minute) between actions.

 Once generated, BLOC strings are be tokenized (Fig.~\ref{fig:bloc_process}b) into bi-grams (two-letter words), then we can represent any social media account as a vector of BLOC words. A collection of point vectors corresponding to multiple accounts make up the BLOC matrix in Fig.~\ref{fig:bloc_process}c.
In this vector space model, each account is represented as a point $(w_1, w_2,...,w_k)$ in $k$-dimensional vector space where each dimension $i$ corresponds to a BLOC word. The weight $w_i$ represents how well an account is described by word $i$. For each account $a$, we instantiated $w_i$ with the TF-IDF weight~\cite{TFIDF}, the product of the term frequency (TF) and the inverse document frequency (IDF):
\begin{equation}
  w_i(a) = f_i(a) \left(1 + \log \frac{D}{d_i} \right)
\end{equation}
where $d_i$ is the number of accounts with word $i$ and $D$ is the total number of accounts.

\paragraph{\textbf{Botometer}}

Botometer,\footnote{\url{botometer.org}} is a publicly available supervised machine learning system that classifies a given Twitter account as bot or human-controlled. Since the release of the first version in 2016, Botometer~\cite{davis2016botornot} (formerly BotOrNot) has been cited over 2,000 times and used extensively by research published across diverse venues including Science~\cite{vosoughi2018spread} and Nature~\cite{pennycook2021shifting}. Like any Machine-Learning model, Botometer is not perfect, but with an F1 of 0.77~\cite{sayyadiharikandeh2020detection} which was calculated from a heterogeneous dataset, it remains one of the most robust methods for bot detection.

Botometer-V4 (the current version of Botometer at the time of writing) \cite{sayyadiharikandeh2020detection} utilizes over 1,000 features that can be grouped into six categories that focus on different account characteristics including, metadata from the accounts (\textit{e.g.}, numbers of friends and followers), retweet and mention networks, temporal features (\textit{e.g.}, frequency of posts), content information, and sentiment. In the deployed system, different classifiers in an ensemble are trained on different accounts types, and then these classifiers vote to obtain the final bot score~\cite{sayyadiharikandeh2020detection}. Here however, we only utilize the representation power of Botometer by extracting its features which serve as input to BNN and DNN.

\subsection{\textbf{Bayesian Neural Networks}}\label{subsec:bnn}
BNNs are extensions of traditional DNNs, aiming to optimize distributions of weights at each layer.\footnote{In contrast to their deterministic counterparts, which define a fixed set of weights over the model.} The aim of BNNs is to approximate a posterior distribution over the weights, given a dataset $q(\bold{W}|D)$. This allows predictions of quantities through a posterior distribution $q(\bold{y}|\bold{x},D)$, integrated over the weights. The formulation of such a posterior is intractable and therefore one must turn to Bayesian inference techniques during optimization. Traditional approaches define the posterior distribution to be a fully factorized Gaussian (diagonal covariance) $q(\bold{W})$, such that the evidence-lower bound (ELBO) can be maximized between the approximated posterior and the prior. Note that maximizing the ELBO is equivalent to minimizing the KL Divergence. This assumption, although more advantageous, tends to be limiting in terms of learned network complexity. Another approach provided in Louizos and Welling \cite{pmlr-v70-louizos17a}, is to instead approximate the posterior as a product of fully factorized Gaussian and a mixing density, Eq. \ref{eq:mixing}, where $q(\bold{z})$ is a vector of random variables. The random variables act multiplicatively on the means of the mixing density to reduce computational complexity.

\begin{equation}\label{eq:mixing}
    q(\bold{W}) = \int q(\bold{W} | \bold{z}) q(\bold{z}) d\bold{z}
\end{equation}

The result is a more flexible posterior distribution over the weights, capable of learning multi-modal dependencies between weights. However, this too becomes intractable, and therefore, an approximate lower bound of entropy must be constructed. This can be done through an auxiliary distribution $r(\bold{z}|\bold{W})$, equivalent to performing variational inference on an augmented probability space \cite{pmlr-v70-louizos17a}. Moreover, the approximated posterior remains true given the fact that the auxiliary distribution can be marginalized out. As stated in Louizos and Welling \cite{pmlr-v70-louizos17a}, the tightness of the bound on $q(\bold{W})$ (and therefore the quality of it) directly depends on the ability of $r(\bold{z}|\bold{W})$ to approximate the posterior of $q(\bold{z} | \bold{W})$, and is therefore chosen to be represented with inverse normalizing flows. The choice of normalizing flows allows analytic computation of the marginals through bijective transformations. The approximate posterior is then constrained through Eq. \ref{eq:KL} during training, acting as a regularization term in conjunction with traditional loss functions.

\begin{align}\label{eq:KL}
    \mathcal{L}_{KL.} &= -KL(q(\bold{W})\|p(\bold{W})) \nonumber \\
    &= \mathbb{E}_{q(\bold{W},\bold{z}_T)}[-KL(q(\bold{W}|\bold{z}_{T_f}) \| p(\bold{W})) \nonumber \\
    & + \log r(\bold{z}_{T_f} | \bold{W}) - \log q(\bold{z}_{T_f})]
\end{align}

We also extend our network to capture the aleatoric uncertainty component using the methodology described in Kendal et al. \cite{kendall2017uncertainties}. For each input the network produces a latent variable $\bold{f}$, along with with the aleatoric uncertainty component $\bold{s} = \log \sigma$. We choose to interpret the network output ($\bold{s}$) as the logarithm of the uncertainty, allowing $\sigma$ to be positive definite through an exponential transformation $\sigma = e^{\bold{s}}$. We then define a Gaussian distribution over the latent variate such that:

\begin{equation}\label{eq:monte_dist}
    \begin{aligned}
        \bold{\hat{f}} | \bold{W} & \sim N(\bold{f}_{\bold{W}}, \sigma_{\bold{W}}) \\
        p & = \text{Sigmoid}(\bold{\hat{f}})
    \end{aligned}
\end{equation}
where $\bold{W}$ are the weights of the network. The expected log-likelihood, and therefore loss function is given by Eq. \ref{eq:aleatoric_loss}, where the subscript c denotes the associated class.

\begin{equation}\label{eq:aleatoric_loss}
    \log \mathbb{E}_{N(\bold{f}, \sigma)} [p_c]
\end{equation}

As mentioned in Kendal et al. \cite{kendall2017uncertainties}, it is not possible to integrate out the Gaussian distribution, and as such Monte Carlo integration must be deployed. At training time this amounts to and extra sampling step to draw samples following Eq. \ref{eq:sampling} in which we take the expected value of the latent variable to produce our probability.

\begin{equation}\label{eq:sampling}
    \bold{\hat{f}} = \bold{f}_{\bold{W}} + \sigma_{\bold{W}} \cdot \epsilon \, , \; \; \; \; \; \epsilon \sim N(0,1)
\end{equation}

\begin{figure}[!]
    \centering
    \includegraphics[width=0.8\textwidth]{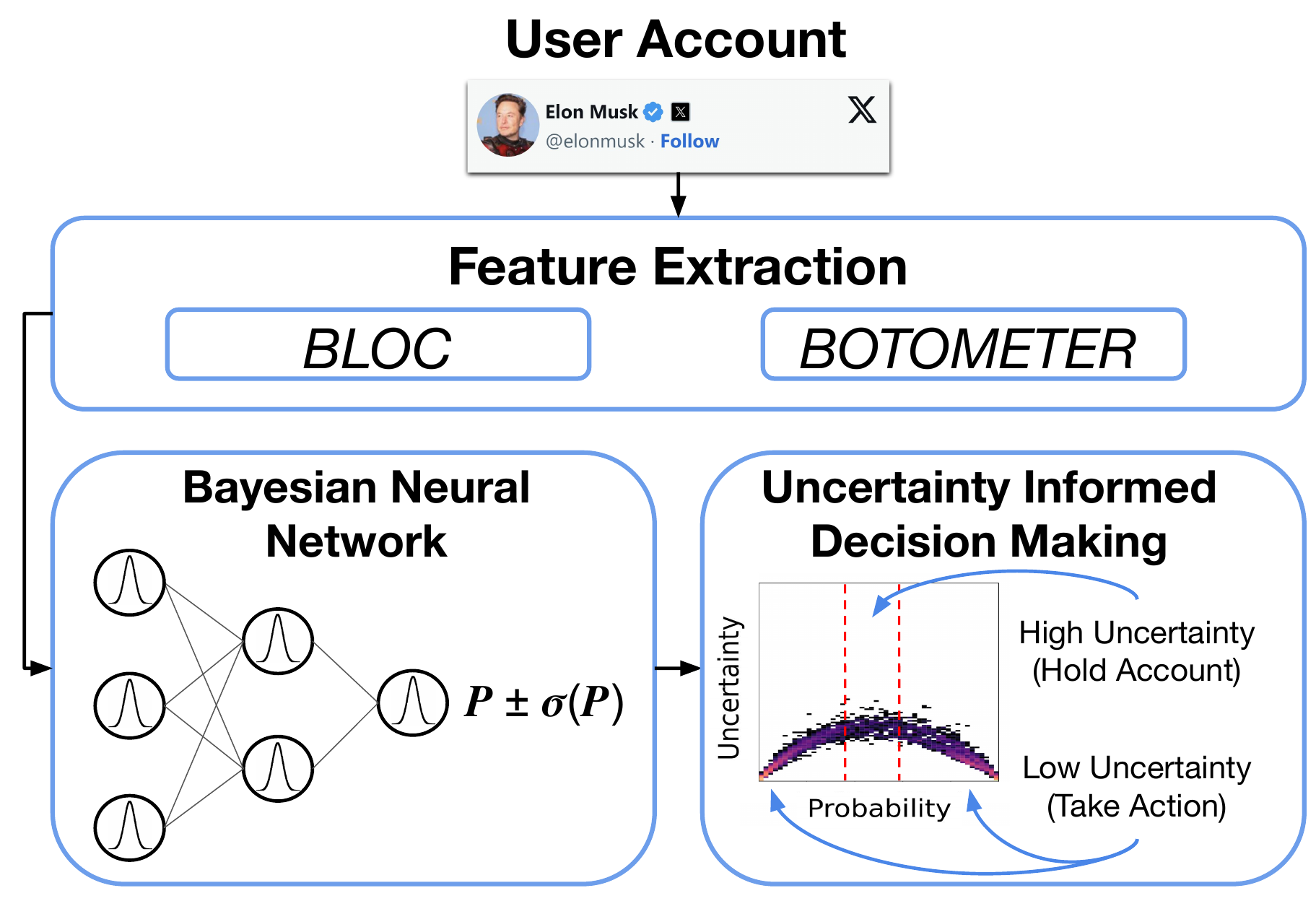}
    \caption{\textbf{Analysis Pipeline:}
    Schematic representation of uncertainty aware decision making in bot detection. The Bayesian Neural Network (BNN) structure is characterized by Multiplicative Normalizing Flows (MNF) \cite{pmlr-v70-louizos17a}, batch normalization, and SELU activation functions \cite{klambauer2017self}. The output of the network is the probability of a bot account, along with the epistemic and aleatoric uncertainties. These uncertainties can be combined in quadrature.
    }
    \label{fig:analysis_pipeline}
\end{figure}

We inherit the design philosophy from Fanelli and Giroux \cite{fanelli2024eluquant}, in which we first design a minimal complexity DNN as the basis for our BNN. Bayesian blocks characterized by Multiplicative Normalizing Flows (MNF) layers \cite{pmlr-v70-louizos17a} are utilized at each layer. The analysis pipeline is shown in Fig. \ref{fig:analysis_pipeline}.

As stated in Fanelli and Giroux \cite{fanelli2024eluquant}, SELU activation functions, as presented by Klambauer et al. \cite{klambauer2017self}, possess inherent self-normalizing properties, which ensure non-vanishing gradients. Their self-normalization nature could provide cases in which batch normalization is not needed, although this is data-dependent. We utilize SELU along with batch normalization to improve network convergence \cite{klambauer2017self}.
The output of the network provides a probability of the account being a bot given a BLOC input; a representation of both the accounts actions and content. The task is treated as a binary classification loss where we deploy Binary Cross-Entropy, Eq. \ref{eq:bce}, coupled to the KL term in Eq. \ref{eq:KL}.

\begin{equation}\label{eq:bce}
    \mathcal{L}_{BCE.} = - \frac{1}{N} \sum_i ^ N y_i \cdot \log(p(y_i)) + (1 - y_i) \cdot \log(1 - p(y_i))
\end{equation}

The resulting loss function is the given by Eq. \ref{eq:loss}, where $\alpha$ is a scaling parameter to allow the contribution of the BCE term. Trivial optimization techniques showed values on the order of $\alpha \sim 10^{-4}$ to be suitable given the relative scales. 

\begin{equation}\label{eq:loss}
    \mathcal{L} = \mathcal{L}_{BCE.} + \alpha \mathcal{L}_{KL.}
\end{equation}

\section{Datasets and Experimental Setup}\label{sec:exps}

\textcolor{black}{
For this study, we utilized a set of datasets from the Bot Repository,\footnote{\url{https://botometer.osome.iu.edu/bot-repository}} which consists of labeled Twitter account data gathered by various researchers between 2017 and 2019. These datasets were specifically created to assist in the development of bot detection models and encompass accounts from diverse domains, including political discourse, celebrity interactions, and general social media activity. In total, the dataset includes over 10 million tweets. The datasets used in this study, along with the number of labeled bot and human accounts, are presented in Table \ref{tab:bot_human_counts}.}

\begin{table}[h]
\centering
\begin{tabular}{lrr}
\toprule
\textbf{Source} & \textbf{Bot Counts} & \textbf{Human Counts} \\
\midrule
midterm-18 \cite{yang2020scalable}            & 0     & 7458  \\
cresci-stock-18 \cite{cresci2018fake}       & 7102  & 6172  \\
pronbots-19 \cite{yang2019arming}           & 17884 & 0     \\
botometer-feedback-19 \cite{yang2019arming}  & 139   & 379   \\
cresci-17 \cite{cresci2017paradigm}             & 7049  & 2760  \\
celebrity-19 \cite{yang2019arming}    & 0     & 20549 \\
gilani-17 \cite{gilani2017bots}             & 914   & 1576  \\
verified-19 \cite{yang2020scalable}           & 0     & 1891  \\
astroturf-20 \cite{sayyadiharikandeh2020detection}              & 502   & 0     \\
vendor-purchased-19 \cite{yang2019arming}   & 1069  & 17    \\
varol-17 \cite{varol2017online}              & 732   & 1496  \\
political-bots-19 \cite{yang2019arming}     & 62    & 0     \\
cresci-rtbust-19 \cite{mazza2019rtbust}       & 352   & 340   \\
botwiki-19 \cite{yang2020scalable}            & 691   & 0     \\
\midrule
\textbf{Total}         & 36496 & 42638 \\
\bottomrule
\end{tabular}
\caption{\textbf{Bot and Human Distribution across Source Datasets:}
Account distribution across source datasets used within this study. The accounts are extracted from individual datasets, and then combined to form our final dataset.
}
\label{tab:bot_human_counts}
\end{table}


The BNN is trained using a traditional $70/15/15 \%$ split, in which we make the distribution of humans and bots equal prior to splitting. This removes any bias the network may incur due to class imbalance. This sampling results in an excess of human accounts which we use as additional performance measures. The classwise distribution of the following subsets can be found in Table. \ref{tab:subsets}.

\begin{table}[h]
    \centering
    \begin{tabular}{c c c}
    \toprule
        Subset & Bots  & Humans  \\
    \midrule
        Training      & 25566 & 25528 \\
        Validation    & 5443  & 5506  \\
        Testing       & 5487  & 5462 \\
        Excess  & 0 & 6142 \\
        \bottomrule
    \end{tabular}
    \caption{\textbf{Distribution of Accounts:}
    The distribution of account types across the training, validation, testing and excess datasets. The datasets used at training are split such that the number of human and bot accounts are approximately equal, removing potential biases towards a singular class. 
    }
    \label{tab:subsets}
\end{table}

The Adam optimizer is used, along with a Cosine Annealing learning rate scheduler with an initial learning rate of $5 \times 10^{-4}$. We deploy early stopping, defining the convergence when the validation loss is no longer decreasing after five epochs. The number of epochs for early stopping is chosen to reflect the stability of the training and account for fluctuations. Information regarding training is summarized in Table. \ref{tab:training}. Note that the DNN is trained under the same conditions for fair comparison, modulo certain irrelevant components such as the computation of samples for learning the aleatoric uncertainty from data.

\begin{table}[h]
\centering
\begin{tabular}{ c  c } 
\hline
\textbf{Training Parameter} & \textbf{value} \\
\hline
Aleatoric Samples & 1k \\
Batch Size & 1024 \\
Training GPU Memory & $\sim 2$GB \\
Trainable Parameters & 91,777 \\
Initial Learning Rate & $5 \times 10^{-4}$ \\
Maximum Epochs & 100 \\
KL Scale & $10^{-4}$ \\
Wall Time & $\sim$ 2 minutes \\
Network memory on local storage & $\sim 8$ MB \\
\hline \\
\end{tabular}
\caption{\textbf{Training:}
Summary of training parameters and computational usage. Training is performed with an Intel i9-14900KF CPU, Nvidia RTX 4090 24GB GPU and 64GB of RAM.
}
 \label{tab:training}
 \end{table}

At inference, we sample a set of ten thousand weights from the network posterior for each account. This, in turn, provides a posterior distribution of the predicted probability of being a bot account. We then take the expected value (the mean) as the final probability and compute the standard deviation on this distribution to provide the epistemic (model) uncertainty. The aleatoric uncertainty is taken to be the average. In this case, we are assuming a Gaussian uncertainty profile on the output, which is a good approximation given the choice of Gaussian prior. Inference statistics can be found in Table. \ref{tab:inference}.

\begin{table}[h]
\centering
\begin{tabular}{ c  c } 
\hline
\textbf{Inference Parameter} & \textbf{value} \\
\hline
Number of Samples (N) & 10k \\
Batch Size (BLOC) & 75 \\
Batch Size (Botometer) & 35 \\
Inference GPU Memory & $\sim 17$ GB \\
Inference Time per Event & $\sim 8ms$ \\
\hline \\
\end{tabular}
\caption{\textbf{Inference:}
Specification of performance at inference. Inference is performed with an Intel i9-14900KF CPU, Nvidia RTX 4090 24GB GPU and 64GB of RAM.
}
 \label{tab:inference}
 \end{table}


\section{Results}\label{sec:results}

In this section, we discuss the results of predicting the labels (\textit{bot} or \textit{human}) of accounts with our network, a BNN inspired by the Event-Level-Uncertainty Quantification (ELUQuant) \cite{fanelli2024eluquant} work originally developed for nuclear physics. This approach allows for the calculation of aleatoric and epistemic uncertainty in the predictions of whether Twitter/X accounts are bots or humans.

\begin{figure}[h]
    \centering
    \begin{subfigure}[b]{0.48\textwidth}
        \centering
        \includegraphics[width=\textwidth,trim=0 0 0 1.2cm, clip]{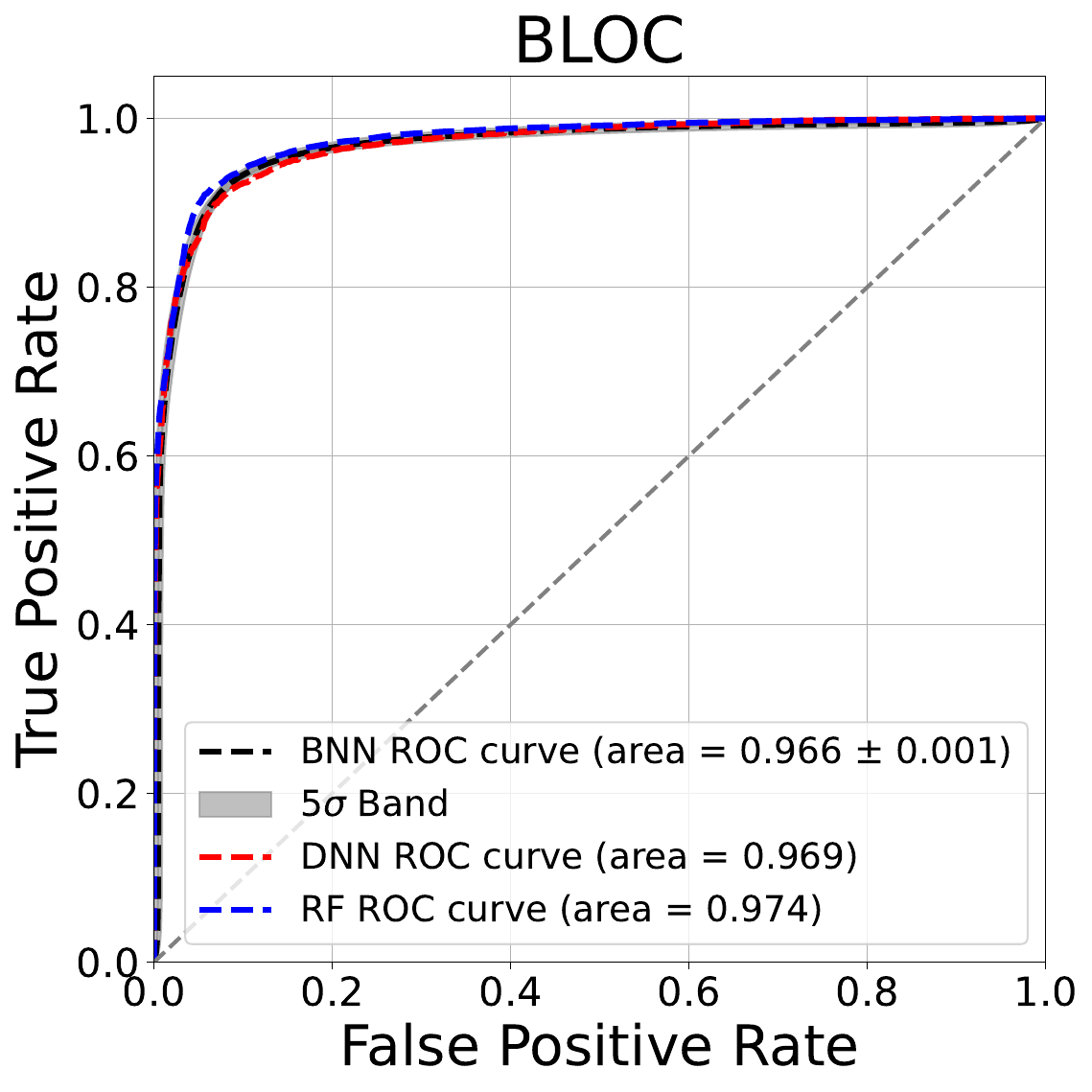}
        \caption{BLOC}
    \end{subfigure}
    \hfill
    \begin{subfigure}[b]{0.48\textwidth}
        \centering
        \includegraphics[width=\textwidth,trim=0 0 0 1.2cm, clip]{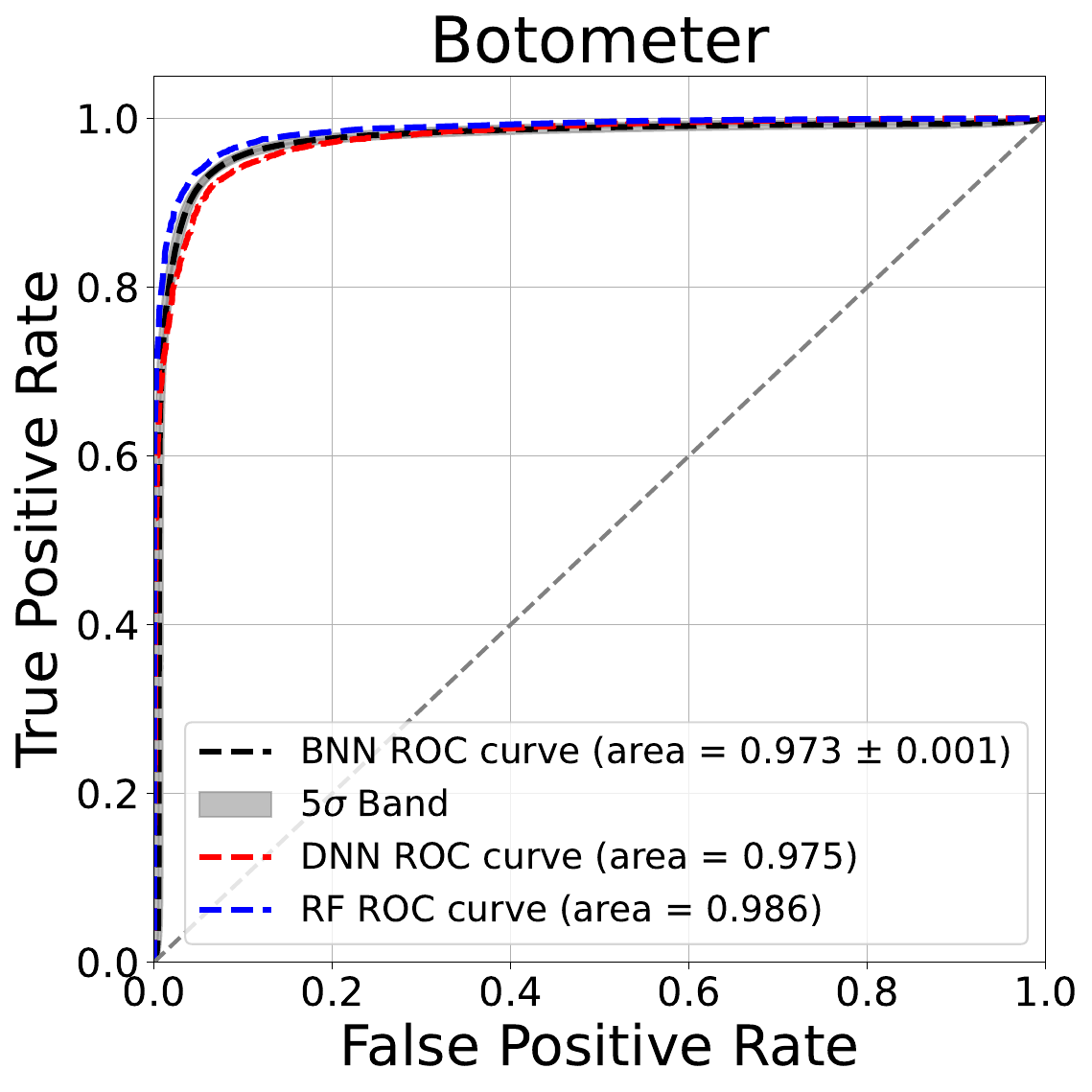}
        \caption{Botometer}
    \end{subfigure}
    
    \caption{\textbf{Overlayed ROC Curves:}
    Receiver Operating Characteristic (ROC) Curves for the Bayesian Neural Network (BNN), Deep Neural Network (DNN) and Random Forest (RF), trained on BLOC features (a) and Botometer features (b). The uncertainty band on the BNN curves is obtained through a bootstrapping method, in which we sample the posterior over the weights to obtain uncertainties on the False Positive Rate (FPR) and True Positive Rate (TPR) at each threshold. Note the DNN and BNN perform consistently within error. RF outperforms the networks due to its increased ability to operate datasets with lower statistics more efficiently.
    }
    \label{fig:roc_comparison}
\end{figure}

It is also important to compare the performance of the BNN with its deterministic counterpart, the Deep Neural Network (DNN). A BNN should perform at least as well as its deterministic counterpart.
This has been demonstrated in the following way.
We evaluate our model using standard methods, namely the Receiver Operating Characteristic (ROC) Curve and the associated Area Under the Curve (AUC). These metrics allow us to avoid making hard threshold cuts in the probability space and reflect a model's performance across various thresholds. Thus, AUC is an ideal metric for such use cases. We then compare the results to the deterministic counterpart of our BNN, a DNN, along with the Random Forest (RF) from Nwala et al. \cite{nwala2023language}. We extract both BLOC and Botometer features for the same set of users.
Figure \ref{fig:roc_comparison} shows a comparison between three methods, with the AUC indicated in the legend. The left plot contains the ROC curves for the algorithms trained on BLOC features, where the error is calculated through bootstrapping over the posterior on the probability. This approach allows us to produce $5\sigma$ bands over both the True Positive Rate (TPR) and False Positive Rate (FPR). The right plot provides the same ROC Curves for algorithms trained on Botometer features. We note an AUC for the BNN of $0.966 \pm 0.001$, which agrees with the deterministic DNN that achieves an AUC of $0.969$ on BLOC features. The same agreement between the BNN and DNN is also seen in Botometer features with the BNN obtaining an AUC of $0.973 \pm 0.001$ and the DNN obtaining an AUC of $0.975$. In both cases, the RF outperforms the DNN and BNN due to its ability to operate more efficiently with smaller training sample sizes.

\begin{figure}[!b]
    \centering
    \begin{subfigure}[b]{\textwidth}
        \centering
        \includegraphics[width=0.328\textwidth]{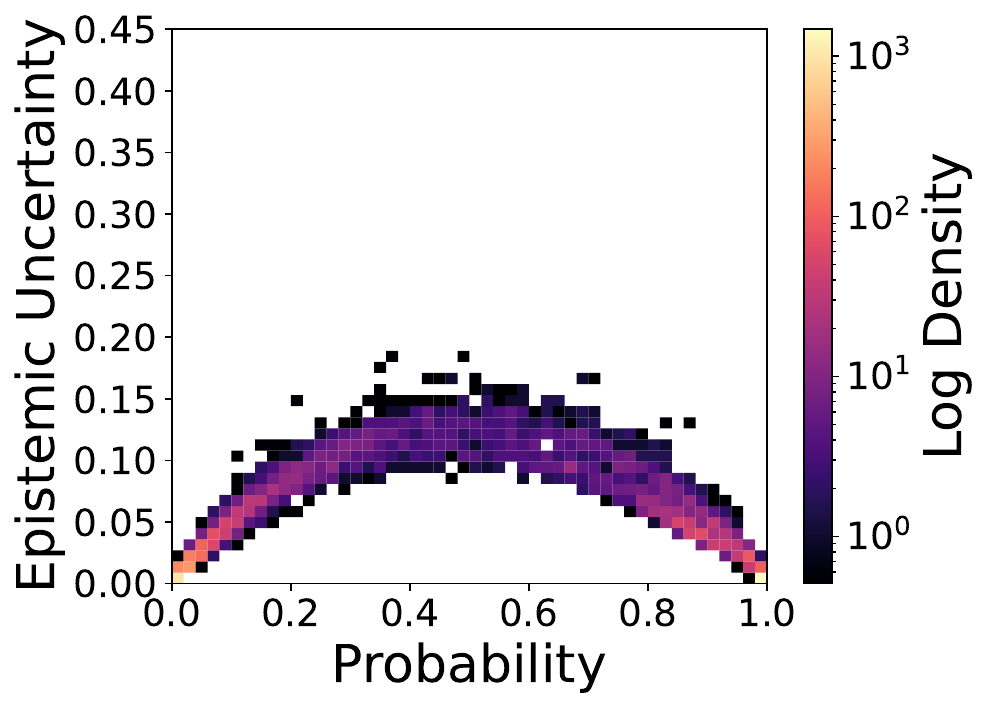} %
        \includegraphics[width=0.328\textwidth]{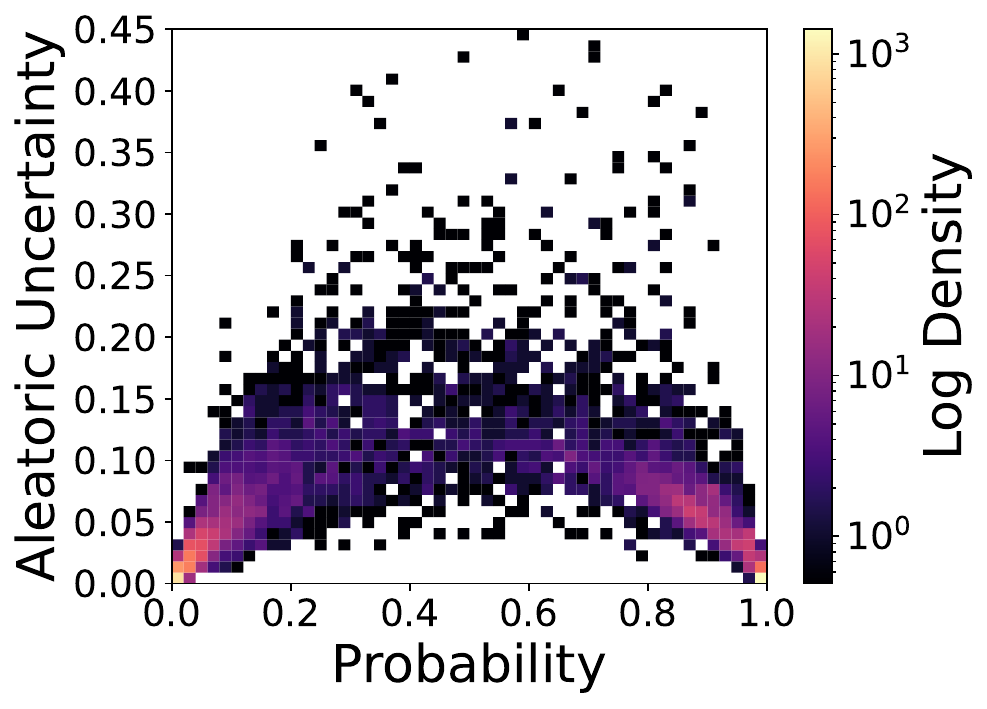} %
        \includegraphics[width=0.328\textwidth]{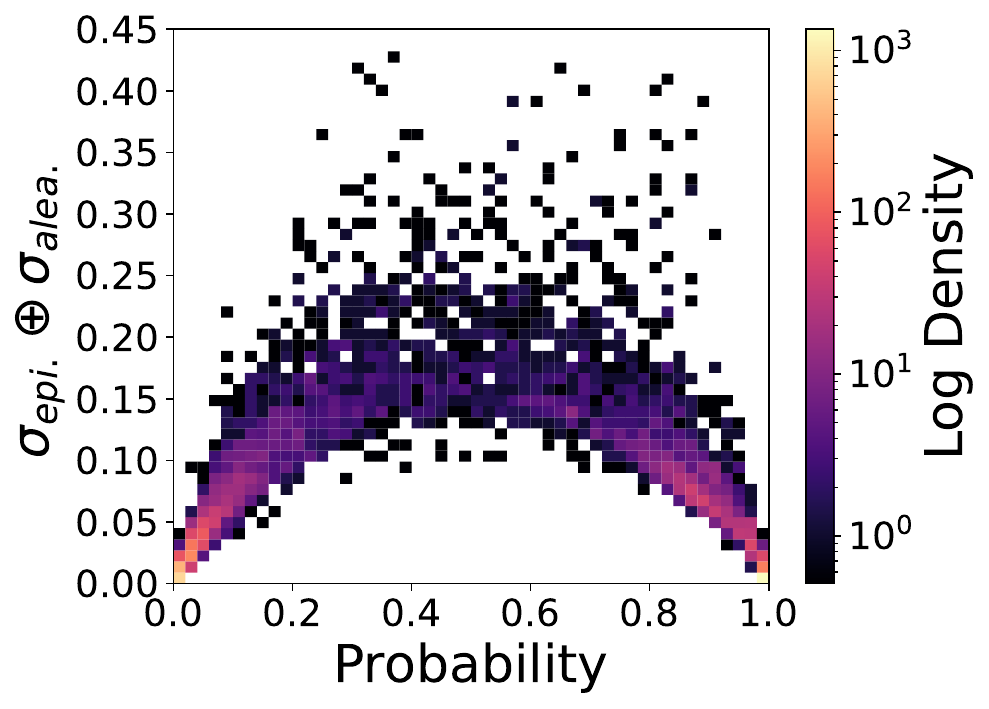} 
        \caption{BLOC}
    \end{subfigure}

    \begin{subfigure}[b]{\textwidth}
        \centering
        \includegraphics[width=0.328\textwidth]{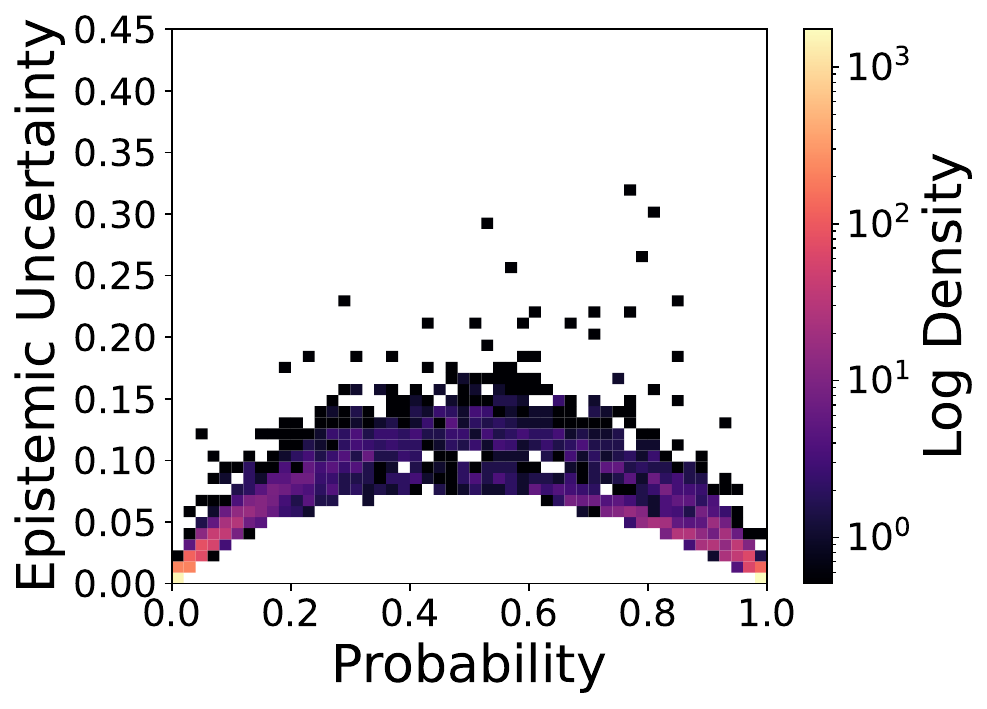} %
        \includegraphics[width=0.328\textwidth]{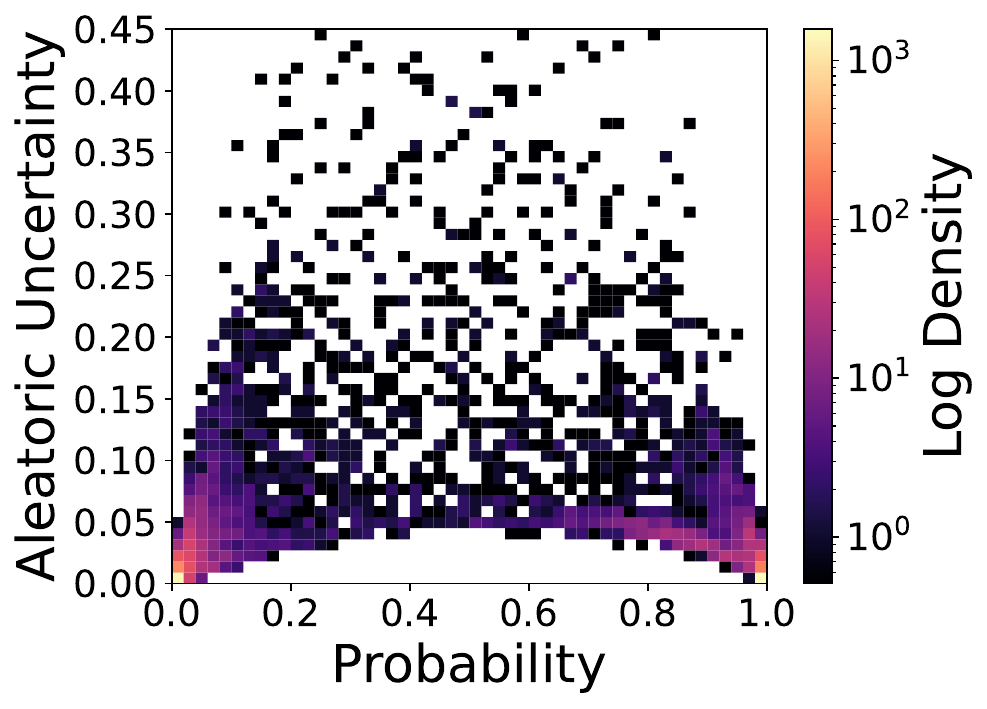} %
        \includegraphics[width=0.328\textwidth]{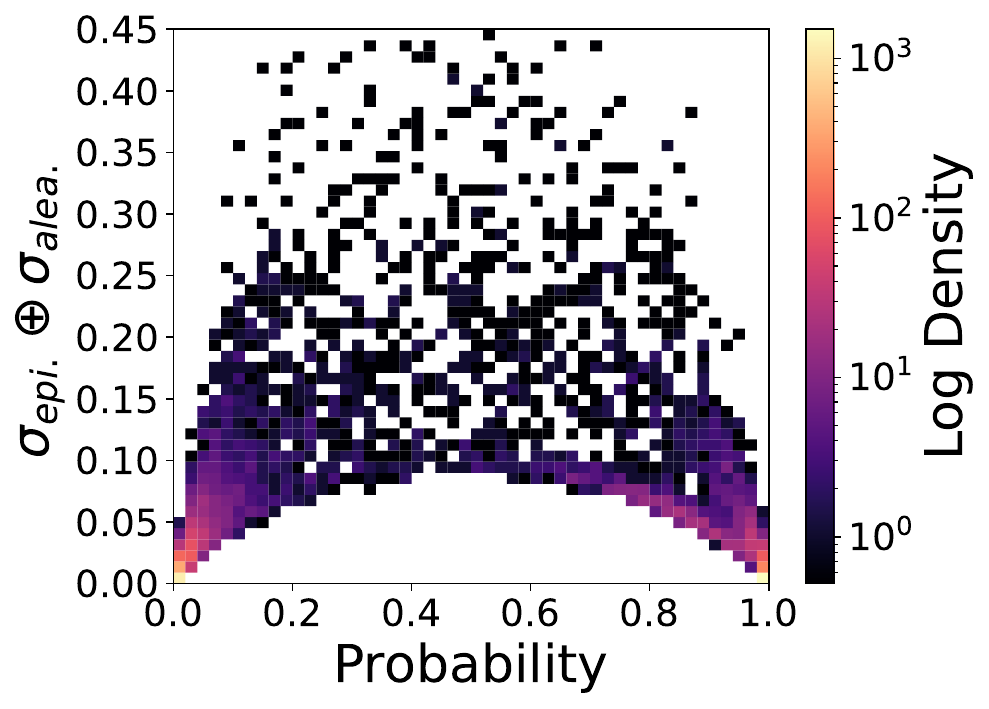} 
        \caption{Botometer}
    \end{subfigure}

    \caption{\textbf{Uncertainty as a Function of Probability:}
    Epistemic uncertainty, aleatoric uncertainty and the two in quadrature as function of model probability for the models trained on (a) BLOC features (top row) and (b) Botometer features (bottom row). Note the parabolic like shape of the epistemic distribution, with maximum uncertainty around the decision boundary ($ < p_{bot} > = 0.5$). For a well calibrated Bayesian model, this is the expected behavior of the epistemic uncertainty. The aleatoric uncertainty is dictated by the available data, and therefore there exists no expectation on its distribution. The two uncertainties in quadrature produce a convolution of the two, epistemic and aleatoric.
    }
    \label{fig:uncertainty}
\end{figure}

We also aimed to validate the epistemic uncertainty produced by the network, where we expect maximum uncertainty for probability values $ P(\text{bot})  \sim 0.5$.
Figure \ref{fig:uncertainty} reports the aleatoric uncertainty, epistemic uncertainty, and the sum in quadrature for the total uncertainty.

One can notice that, in general, the uncertainty is larger when the prediction is more ambiguous, that is, around a probability of 0.5, and it is smaller when it is close to 0 (account identified as human) or 1 (account identified as a bot).
The reader should be reminded that the uncertainties are provided at the Twitter/X account level, meaning that our BNN provides an output probability of being a bot account along with the associated uncertainties, both aleatoric and epistemic.
Another observation is that the aleatoric uncertainty, which captures the randomness in our data and its propagation in the network's predictions, is generally more spread than the epistemic uncertainty and can reach higher values.

\begin{figure}[!]
    \centering
    \begin{subfigure}[b]{0.45\textwidth}
        \centering
        \includegraphics[width=\textwidth]{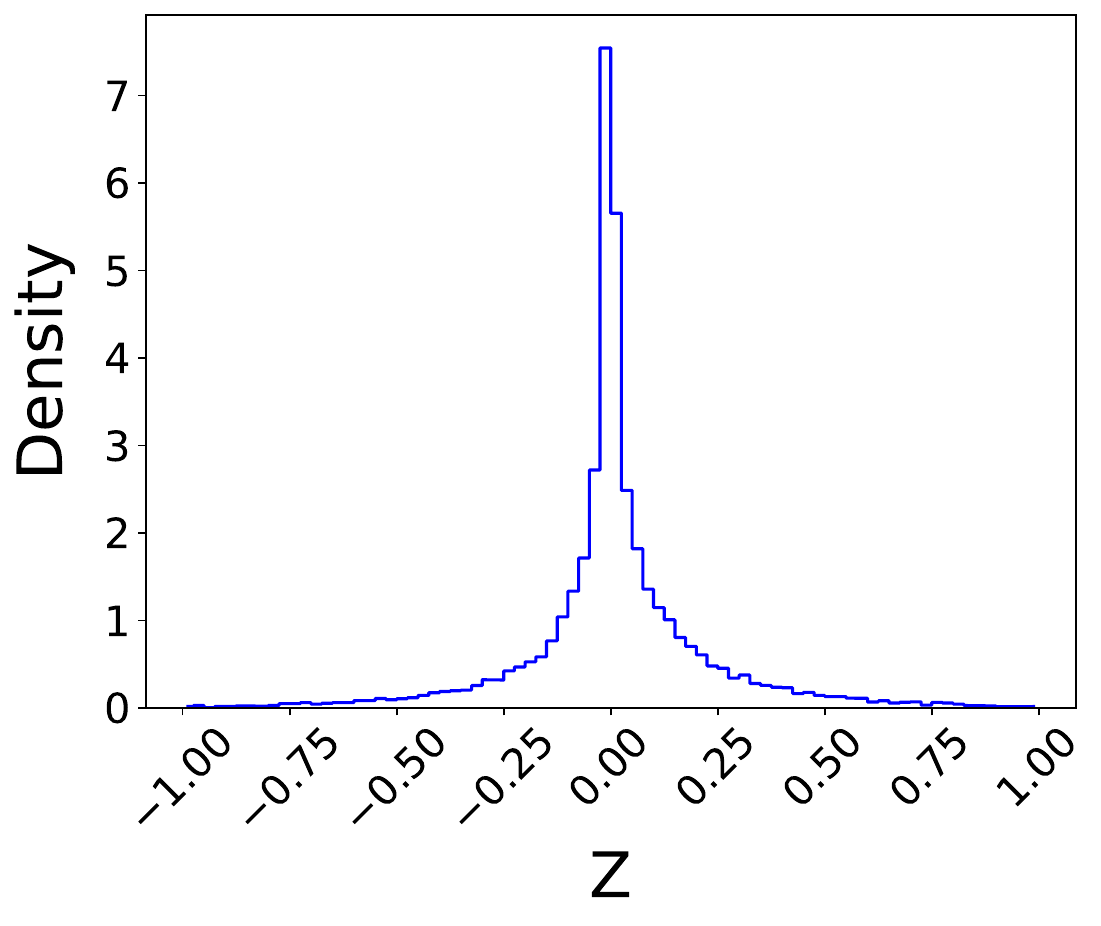}
        \caption{BLOC}
        \label{fig:bloc_epi}
    \end{subfigure}
    \hfill
    \begin{subfigure}[b]{0.45\textwidth}
        \centering
        \includegraphics[width=\textwidth]{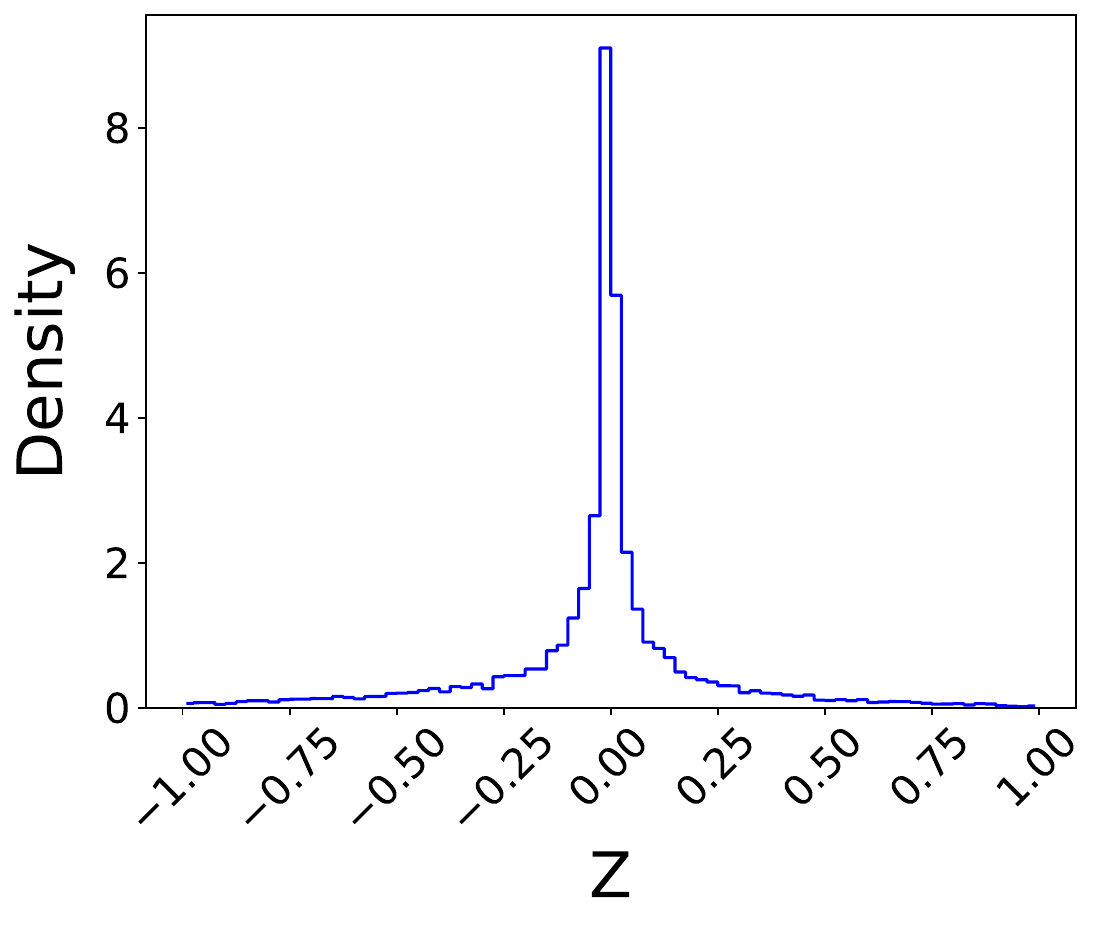}
        \caption{Botometer}
        \label{fig:botometer_epi}
    \end{subfigure}
    
    \caption{\textbf{Z-Score Tests for Decoupling Aleatoric and Epistemic Uncertainties}:
    Z-Score tests for the decoupling of aleatoric and epistemic uncertainties, for BLOC (a) and Botometer features (b). The results support the ability of the network to decouple epistemic and aleatoric uncertainty during training, in which the majority of samples lie within $|Z| < 0.5$.
    }
    \label{fig:z_score}
\end{figure}

We consolidated the uncertainty quantification by running a closure test, consisting of training the network (i) without including the aleatoric term, thereby predicting only the epistemic uncertainty, and (ii) including both aleatoric and epistemic terms. Details on how these two scenarios can be implemented are discussed in the Sec. \ref{sec:methods} and are also described in \cite{fanelli2024eluquant}. 
We demonstrated through a Z-score test, visualized in Fig. \ref{fig:z_score}, that the epistemic uncertainties obtained on the accounts (considering the results from methods (i) and (ii) at the account level) produce consistent epistemic uncertainties (\textcolor{black}{the majority} of values are within $|\text{Z}| \leq 0.5$). This closure test supports the fact that the quantified aleatoric uncertainty is decoupled from the epistemic uncertainty, with the latter appearing, on average, different from the former.

After consolidating our results, as discussed, we compare the performance of our network with respect to state-of-the-art works utilizing BLOC features \cite{nwala2023language} and Botometer \cite{yang2022botometer} with random forest (RF) as a classifier, as done in those papers.
We evaluate the performance using precision, recall, and F1 metrics. Our method outperforms other methods in terms of recall and F1, and is nearly on par in terms of precision, as shown in Table \ref{tab:table_performance}. \textcolor{black}{We use a threshold value of 0.5, representing the natural decision boundary of a Sigmoid function.}

\begin{table}[]
    \centering
    \scalebox{0.85}{
    \begin{tabular}{c c c c c c c c}
    \toprule
    &  & \multicolumn{3}{c}{\textbf{BLOC}} & \multicolumn{3}{c}{\textbf{Botometer}} \\
     \cmidrule(lr){3-5} \cmidrule(lr){6-8}
     Subset & Model & Precision & Recall & F1 & Precision & Recall & F1 \\
     \midrule
     Human (Test Split) 
        &  BNN & \textbf{0.928 $\pm$ 0.003} & 0.921 $\pm$ 0.004 & 0.924 $\pm$ 0.004  & \textbf{0.938 $\pm$ 0.003} & 0.939 $\pm$ 0.003 & 0.939 $\pm$ 0.003 \\
        &  DNN & 0.919 $\pm$ 0.004 & 0.906 $\pm$ 0.004 & 0.912 $\pm$ 0.004 & 0.927 $\pm$ 0.004 & 0.926 $\pm$ 0.004 & 0.926  $\pm$ 0.004\\
        &  RF \cite{nwala2022general} & 0.916 $\pm$ 0.004 & \textbf{0.937 $\pm$ 0.003} & \textbf{0.927 $\pm$ 0.004} & \textbf{0.938 $\pm$ 0.003} & \textbf{0.950 $\pm$ 0.003} & \textbf{0.944 $\pm$ 0.003} \\
     \midrule
     Bot (Test Split)   
        &  BNN & 0.922 $\pm$ 0.004 & \textbf{0.929 $\pm$ 0.003} & \textbf{0.925 $\pm$ 0.004} & 0.939 $\pm$ 0.003 & \textbf{0.938 $\pm$ 0.003} & 0.939 $\pm$ 0.003 \\
        &  DNN & 0.908 $\pm$ 0.004 & 0.920 $\pm$ 0.004 & 0.914 $\pm$ 0.004 & 0.926 $\pm$ 0.004  & 0.927 $\pm$ 0.004 & 0.927  $\pm$ 0.004 \\
        &  RF \cite{nwala2022general}  & \textbf{0.936 $\pm$ 0.003 } & 0.915 $\pm$ 0.004 & \textbf{0.925 $\pm$ 0.004} & \textbf{0.949 $\pm$ 0.003} & \textbf{0.938 $\pm$ 0.003} & \textbf{0.943 $\pm$ 0.003} \\
     \midrule
     Excess Human 
        &  BNN & N/A & 0.920 $\pm$ 0.003 & N/A & N/A  & 0.939 $\pm$ 0.003 & N/A \\
        &  DNN & N/A  & 0.438 $\pm$ 0.006 & N/A & N/A  & 0.501 $\pm$ 0.006 & N/A \\
        &  RF \cite{nwala2022general} & N/A  & \textbf{0.938 $\pm$ 0.003} & N/A & N/A  & \textbf{0.952 $\pm$ 0.003} & N/A \\
      \bottomrule
    \end{tabular}}
    \caption{\textbf{Performance Comparison of Classifiers Using BLOC and Botometer Features:}
    Bayesian Neural Network (BNN) used in our work, which is compared to a Deep Neural Network (DNN) and to a Random Forest (RF) with features from the Behavioral Language for Online Classification (BLOC) and Botometer. Precision, recall and F1 have been computed for human accounts and bot accounts (using the test dataset which is a mixture of human and bot accounts), and using other bot accounts only not present in our test dataset to test generalization.
    }
    \label{tab:table_performance}
\end{table}
We note that the BNN is more generalizable to the excess human accounts contained within the dataset. Since we sample a 50/50\% class split at training, validation, and testing, we retain the excess accounts as an additional measure of performance. Note that this dataset contains only human accounts, and therefore the only meaningful metric that remains is recall. Precision by default will be perfect (1.0) given no potential for false-positives (bots labeled as humans), which in turn will effect the F1 score. We report these values as N/A to not introduce confusion. The BNN is more able to capture a generalized weight distribution in the form of a posterior in comparison to the DNN. Note that both models have been regularized in the same manner apart from the inherent Kullback-Leibler (KL) divergence term appearing for the BNN, which controls the distribution of the learned weights under the Gaussian prior.
After verifying that our performance is on par with or even surpasses other state-of-the-art approaches, we finally utilize the additional information from uncertainty quantification. We show that through uncertainty informed decision making, we are able to surpass performance of both the DNN and RF consistently. This is the novel contribution of our work compared to other works in the field of bot detection.
In the following Table \ref{tab:table_performance_uncertainty}, we show the results obtained by applying a 3$\sigma$ cut based on the quantified uncertainty at the account level. Specifically, we ensure that the predicted outcome is not consistent with a probability of 0.5 (indicating the largest uncertainty in classification) by using the predicted value of the probability and a 3$\sigma$ interval. In other words, we classify only those events that satisfy:
\begin{equation}\label{eq:uqcut}
|P_{pred} - 0.5| > 3\sigma(P_{pred}). 
\end{equation}
Equation \eqref{eq:uqcut} is used for $\sigma(P_{\text{pred}}) = \sigma_{\text{epi.}}(P_{\text{pred}})$, $\sigma_{\text{alea.}}(P_{\text{pred}})$ and $\sigma_{\text{tot.}}(P_{\text{pred}})$, representing the cases of epistemic or aleatoric only, and the total uncertainty in quadrature.
The results show an improvement in performance, across all metrics, over the baseline \textcolor{black}{(\textit{i.e.}, the BNN without uncertainty information)} as uncertainty is introduced into the decision making process. We also report the ``rejection'' fraction, which corresponds to the number of account that are held for further information to be acquired before classification. With regard to cuts utilizing only the epistemic component, we expect that under a robustly characterized epistemic uncertainty, withholding accounts with high epistemic uncertainty should induce performance increases being that we remove regions of the feature space where overlap (between human and bot accounts) persists to a high degree, \textit{i.e.}, regions of low confidence or higher uncertainty. Similarly, with regard to the cuts using only the aleatoric component, we expect that withholding accounts with high uncertainty should induce performance increases being that we remove regions of high stochasticity in the feature space,corresponding to potentially unreliable predictions due to lack of information seen at training time, or simply regions that are well defined but have high variance. The usage of these two in quadrature can then further increase performance by addressing both issues in unison at inference.
This additional information from uncertainty allows for more informed decisions, thereby impacting the decision-making process for Twitter/X accounts, as further discussed below.

\begin{table}[ht]
\centering
\scalebox{0.65}{
\begin{tabular}{llccccccccc}
\toprule
& & & \multicolumn{4}{c}{Human} & \multicolumn{4}{c}{Bot} \\
\cmidrule(lr){4-7} \cmidrule(lr){8-11}
Dataset & Uncertainty & Accuracy  & Precision & Recall & F1-Score& Rejection (\%) & Precision & Recall & F1-Score & Rejection (\%) \\
\midrule
BLOC & Baseline & 92.5 $\pm$ 0.3 & 0.928 $\pm$ 0.004 & 0.921 $\pm$ 0.004 & 0.924 $\pm$ 0.004 & 0 & 0.922 $\pm$ 0.004 & 0.929 $\pm$ 0.004 & 0.925 $\pm$ 0.004 & 0 \\
 & Epistemic & 95.8 $\pm$ 0.2 & 0.958 $\pm$ 0.003 & 0.958 $\pm$ 0.003 & 0.958 $\pm$ 0.003 & 10.6 & 0.959 $\pm$ 0.003 & 0.959 $\pm$ 0.003 & 0.959 $\pm$ 0.003 & 12.2 \\
 & Aleatoric & 96.0 $\pm$ 0.2 & 0.959 $\pm$ 0.003 & 0.961 $\pm$ 0.003 & 0.960 $\pm$ 0.003 & 12.2 & 0.962 $\pm$ 0.003 &  0.960 $\pm$ 0.003 & 0.961 $\pm$ 0.003 & 14.0 \\
 & Quadrature & 96.6 $\pm$ 0.2 & 0.964 $\pm$ 0.003 & 0.967 $\pm$ 0.003 & 0.966 $\pm$ 0.003 & 15.5 & 0.968 $\pm$ 0.003 &  0.965 $\pm$ 0.003 & 0.967 $\pm$ 0.003 & 17.6\\
\midrule
 Botometer & Baseline & 93.9 $\pm$ 0.2 & 0.938 $\pm$ 0.003 & 0.939 $\pm$ 0.003 & 0.965 $\pm$ 0.003 & 0 & 0.939 $\pm$ 0.003 & 0.938 $\pm$ 0.003 & 0.939 $\pm$ 0.003 & 0 \\
 & Epistemic  & 96.5 $\pm$ 0.2 & 0.962 $\pm$ 0.003 & 0.968 $\pm$ 0.003 & 0.965 $\pm$ 0.003 & 7.3 & 0.968 $\pm$ 0.003 & 0.962 $\pm$ 0.003 & 0.965 $\pm$ 0.003 & 7.5\\
 & Aleatoric & 96.5 $\pm$ 0.2 & 0.962 $\pm$ 0.003 & 0.968 $\pm$ 0.003 & 0.965 $\pm$ 0.003 & 7.8 & 0.969 $\pm$ 0.003 & 0.963 $\pm$ 0.003 & 0.966 $\pm$ 0.003 & 10.0\\
 & Quadrature & 97.2 $\pm$ 0.2 & 0.968 $\pm$ 0.003 & 0.975 $\pm$ 0.003 & 0.972 $\pm$ 0.003 & 10.3 & 0.976 $\pm$ 0.003 & 0.968 $\pm$ 0.003 & 0.972 $\pm$ 0.003 & 12.1\\
\bottomrule
\end{tabular}
}
    \caption{\textbf{Performance Comparison with $3\sigma$ Uncertainty Thresholds:}
    Evaluation of the uncertainty informed decision making process, in which a $3 \sigma$ cut is applied on the probability around 0.5 to indicate whether decisions should be made about a user account. $3\sigma$ cuts are applied to the epistemic, aleatoric and uncertainties in quadrature. Note the increase in performance as specific accounts are withheld. Extra information from these accounts can be acquired to reduce uncertainty to a desirable threshold, or allow human intervention on more reasonable sample sizes.
    }
    \label{tab:table_performance_uncertainty}
\end{table}

In summary, our approach features a fully Bayesian framework inspired by ELUQuant \cite{fanelli2024eluquant} to classify Twitter/X accounts as bots or humans, assessing its performance against DNNs and RF models. The BNN demonstrates comparable performance to both the DNN and RF models in terms of AUC, while providing additional information in the form of uncertainty, which is crucial for decision-making at the account level. Closure tests affirm the robustness of our uncertainty quantification, in which we are able to decouple the aleatoric and epistemic components. By applying a $3 \sigma$ uncertainty threshold, we observe improved accuracy and F1 scores, highlighting the utility of uncertainty-aware models in bot detection. This approach not only enhances predictive reliability but also provides deeper insights into model behavior. 

\section{Conclusions}\label{sec:conclusions}

Social bots remain a potent instrument malicious agents utilize to spread disinformation and manipulate the public on social media. To tackle the bot problem and mitigate their serious social, political, or economic harms, researchers have developed multiple bot detection algorithms and tools. However, bot detection continues to be a challenging unsolved problem, because bot behaviors are dynamic and heterogeneous (\textit{e.g.}, spam, fake followers, amplifiers), and different training data and detection models capture a subset of these behaviors. 
This means that different detection models could disagree on whether to label the same account as bot or human-controlled, yet they do not produce any uncertainty to indicate how much we should trust their results.

%
We propose the first uncertainty-aware bot detection algorithm that combines bot detection with uncertainty quantification. Our method is agnostic to bot detection features, demonstrated by deploying it with two existing Twitter/X bot detection feature sets: BLOC and Botometer. Our algorithm captures uncertainty arising from randomness in the account feature space (aleatoric uncertainty) and the uncertainty introduced by the bot detection model (epistemic uncertainty). Notably, while every method can introduce epistemic uncertainty, our proposed architecture actively estimates and accounts for this uncertainty, unlike other methods that may ignore it, leading to potentially erroneous decisions. Furthermore, our method demonstrated exceptional performance, matching or surpassing traditional detection techniques.

Crucially, the uncertainty information of our method has multiple applications. First, it could  inform more effective decision making by allowing social media platforms to carry out targeted interventions (\textit{e.g.}, account suspension) for bots when predictions are made with high confidence and caution (\textit{e.g.}, gathering more data) when predictions are uncertain. This could reduce errors associated with mislabeling accounts as bots. Additionally, uncertainty information can indicate anomalous behavior, raising additional flags for accounts exhibiting such patterns.

Our contribution should be framed in the context of end-to-end analysis pipelines using uncertainty at the account level. As we have shown, our design philosophy is agnostic to input, obtaining similar performance on both BLOC and Botometer features. The network itself can easily be adapted to more complex problems and deploy the same uncertainty aware procedures developed within. Specifically, using uncertainty to isolate subsets of accounts where the network has shown to be unreliable. These accounts can be further monitored over a period of time and reevaluated once more information has been obtained, therefore reducing the potential of false account suspension.

\ 

\noindent {\small{\textbf{Acknowledgements}}\\
The authors acknowledge  William  \&  Mary  Research  Computing for providing computational resources and technical support that have contributed to the results reported within this article.




\ 


%

\noindent {\small{\textbf{Abbreviations}}\\
\textbf{AI}: Artificial Intelligence,
\textbf{API}: Application Programming Interface,
\textbf{BLOC}: Behavioral Language for Online
Classification,
\textbf{SWA}: Stochastic Weight Averaging,
\textbf{ELUQuant}: Event-level-Uncertainty Quantification,
\textbf{BNN}: Bayesian Neural Network,
\textbf{DNN}: Deep Neural Network,
\textbf{ROC}: Receiver Operating Characteristic,
\textbf{AUC}: Area Under the Curve,
\textbf{RF}: Random Forest,
\textbf{TPR}: True Positive Rate,
\textbf{FPR}: False Positive Rate,
\textbf{KL}: Kullback-Leibler,
\textbf{MNF}: Multiplicative Normalizing Flows,

\ 

\noindent {\small{\textbf{Availability of data and materials}}\\
The code used for this work is available at \href{https://github.com/wmdataphys/UncertaintyAwareBotDetection}{https://github.com/wmdataphys/UncertaintyAwareBotDetection}. The raw versions of datasets used within this study can be found at \href{https://botometer.osome.iu.edu/bot-repository/}{https://botometer.osome.iu.edu/bot-repository/}.





%


%

\bibliographystyle{plainnat} 
\bibliography{sample}

\end{document}